\documentclass[pre,superscriptaddress]{revtex4}
\usepackage{graphicx}
\usepackage{amssymb,amsthm,amsmath}
\usepackage{hyperref}
\usepackage{subfigure}
\usepackage{amssymb}
\usepackage{amsfonts}
\newcommand{\beq}[0]{\begin{equation}}
\newcommand{\eeq}[0]{\end{equation}}

\newcommand{\bc}{\begin{center}}
\newcommand{\ec}{\end{center}}

\begin{document}

\title{Solitary and Periodic Waves in Collisionless Plasmas: The Adlam-Allen Model 
Revisited}

\author{J. E. Allen}
\affiliation{Mathematical Institute, University of Oxford, Oxford, OX2
  6GG, UK}

\author{D.\,J. Frantzeskakis}
\affiliation{Department of Physics, National and Kapodistrian University of Athens, Panepistimiopolis, Zografos, Athens 15784, Greece}
\author{N.\,I. Karachalios}
\affiliation{Department of Mathematics,\\ Laboratory of Applied Mathematics and Mathematical Modelling,\\ University of the Aegean, Karlovassi, 83200 Samos, Greece}
\author{P.\,G. Kevrekidis}
\affiliation{Department of Mathematics and Statistics, University of
  Massachusetts, Amherst MA 01003-4515, USA}
\affiliation{Mathematical Institute, University of Oxford, Oxford, OX2
  6GG, UK}
\author{V. Koukouloyannis}
\affiliation{Department of Mathematics,\\ Laboratory of Applied Mathematics and Mathematical Modelling,\\ University of the Aegean, Karlovassi, 83200 Samos, Greece}

\begin{abstract}
We consider the Adlam-Allen (AA) system of partial differential equations which, arguably, 
is the first model that was introduced to describe solitary waves in the context of propagation 
of hydrodynamic disturbances in collisionless plasmas. Here, we identify 
the solitary waves of the model
by implementing a dynamical systems approach. The latter suggests
that the model 
also possesses periodic wave solutions --which reduce to the solitary wave in the limiting case of 
infinite period-- as well as rational solutions which are obtained 
herein. In addition, employing a long-wave approximation via a relevant 
multiscale expansion method, we establish the asymptotic 
reduction of the AA system to the Korteweg-de Vries equation. Such a reduction, 
is not only another justification for the above solitary wave dynamics, but also 
may offer additional insights for the emergence of other possible 
plasma waves. Direct numerical simulations are performed for the study of 
multiple solitary waves and their pairwise interactions. 
The stability of solitary waves is discussed in terms of potentially relevant 
criteria, while the robustness of spatially periodic wave solutions is touched 
upon by our numerical experiments. 

\end{abstract}

\maketitle

\section{Introduction}

The fundamental work of Adlam and Allen in 1958 and 1960~\cite{aa1,aa2} constituted one
of the very first examples of models which may exhibit solitary wave dynamics, arguably 
the very first one in the important field of plasma physics. Indeed, this work preceded the
hallmark efforts of Kruskal and Zabusky in 1965~\cite{kz} concerning the study of
solitary waves and their interactions in the context of the famous Fermi-Pasta-Ulam
problem~\cite{fpu} and its connections with the Korteweg-de Vries (KdV)
equation~\cite{kdv}. In addition, the Adlam-Allen (AA) model and its solitary wave was 
introduced earlier than the seminal work of Washimi and Taniuti \cite{tan} who 
showed that the one-dimensional (1D) long-time asymptotic behavior of small-amplitude 
ion-acoustic waves in plasmas is described by the KdV equation. The above 
works paved the way for numerous investigations in the nonlinear physics of plasmas, 
which proved to be a fertile ground for the study of solitary waves and solitons in integrable
and nearly-integrable equations \cite{ir,kono}. In view of the above, the AA model seems to 
have received far less than its share of interest and associated research attention 
within the field of nonlinear waves and solitons, as has been discussed, e.g., in 
Ref.~\cite{allen3}. A recent revisiting of the relevant subject can be
found in Ref.~\cite{allen4}, where the role of the so-called $j \times
B$ force in a collisionless plasma was discussed. 

Our aim in the present work is to restore this. In particular, upon shortly introducing 
the original motivation and formulation of the AA system of partial differential 
equations (PDEs), we start from its reduced --and more tractable for analysis--
form presented in Ref.~\cite{aa2}. First we derive the Lagrangian density of the system, 
as well as the the Hamiltonian density and momentum of the system. Then, we study the 
linear regime and show that the AA model features the linear dispersion relation of the 
improved Boussinesq equation, which is known to describe bidirectional shallow water waves 
\cite{mjarecent}. Proceeding further, we seek traveling waves that can be supported by the model,  
and use techniques from the theory of nonlinear dynamical systems to study the associated 
second-order ordinary differential equation (ODE). We identify exact and analytically expressed 
(as per the original efforts of~\cite{aa1,aa2}) solitary waves, corresponding to homoclinic orbits 
in the associated phase plane of the conservative dynamical system. We find that these waves 
have speeds between once and twice the characteristic Alfv{\'e}n speed~\cite{aa1}. 
This dynamical systems approach also enables the identification of periodic orbits, 
reminiscent of the elliptic function solutions (corresponding to the so-called ``cnoidal waves'') 
of the KdV equation; these periodic orbits correspond to spatially periodic solutions 
of the original AA system. Interestingly, a degenerate case of the relevant ODE, corresponding to 
the case where the traveling wave propagates with exactly the Alfv{\'e}n speed, leads 
to the existence of a rational-type solution. Furthermore, by implementing a suitable 
multiscale asymptotic expansion, based on a long-wave approximation, we show that the original 
AA system can be approximated by the KdV equation, and that --in the small-amplitude limit-- the 
exact solitary waves of the AA system, reduce to the KdV solitons.   

The above strong justification for the potential of solitary wave dynamics motivates us 
to study the interaction of multiple solitary waves of the AA model, by direct numerical 
simulations. When colliding two such waves, we find the interaction between them to be 
nearly (but not completely) elastic with a clearly observable phase shift between the two solitary
waves. While our findings suggest that the AA model is likely not to be completely
integrable, further investigation of the relevant topic is certainly worthwhile.
We also perform a numerical study on perturbations of the identified spatially periodic 
solutions, which suggests their potential robustness. Furthermore, we show that 
in the small-amplitude (KdV) limit the stability of the travelling waves is justified 
by the relevant stability criterion (of a Vakhitov-Kolokolov
type~\cite{vakh}, namely involving the derivative of a conserved
quantity such as here the momentum with respect to the corresponding
Lagrange multiplier, namely here the speed)
for KdV-type equations; see, e.g., \cite{kuz2,PegoWein92,pelin,kuz1} as
well as references therein.
Interestingly, the monotonicity (implying stability) of the relevant
momentum dependence on speed is preserved throughout the interval of
speeds for which the AA solitary waves are physically relevant,
being suggestive of a qualitatively similar result for arbitrary
speeds; nevertheless, admittedly, we can only support this on the basis
of the above criterion for speeds near the  Alfv{\'e}n speed.
In an interesting parallel development, we leverage recent
work on quadratic operator pencils~\cite{todd} to illustrate that an analogous criterion
for Klein-Gordon type equations also exists on the basis of the
Klein-Gordon momentum and its monotonic dependence on the speed.
We have also evaluated the relevant quantity and have shown that the
corresponding sign of the derivative is also one that reflects stability
in the Klein-Gordon case. While the present model is neither of the
KdV, nor of the Klein-Gordon type, 
the fulfillment of these stability criteria, together with the results of direct numerical 
simulations, allows us to conjecture the generic stability of the AA
model
traveling waves. Finally, we comment 
on the case of rational solutions, due to their mathematical interest.

Here, we should make a few remarks regarding the applicability of our results in plasma physics. 
First, we should point out that, although we establish a strong connection of the AA model with 
the KdV equation, which is known to describe nonlinear ion-acoustic waves in plasmas (see, e.g., 
Ref.~\cite{ir}) the waves supported by the AA model are not directly related to ion-acoustic waves.
Indeed, the AA model describes ``hydrodynamic waves'' in quasi-neutral collisionless plasmas 
propagating perpendicular to a magnetic field.
In comparison, e.g., with the classic work of~\cite{tan} on
ion-acoustic waves, our model
bears velocity both along the x- and the y-direction (although both are
$x$-dependent)
and features both magnetic, as well as electric fields, as opposed to
the single-speed and solely electric-field dependence of~\cite{tan}.
Second, regarding the applicability of the AA model, first we note that 
the early work \cite{aa1} was carried out in the early days of fusion research. Thus, 
the waves supported by the AA model -- which is the main theme of this work -- may find 
applications in fusion research, but also in astrophysical observations. In astrophysics, 
such waves may be associated with the Earth`s bow shock \cite{Earth}, coronal mass ejections 
\cite{Geophys} and plasma release experiments \cite{Plasma}.
More practically, at a physical level, the AA model describes the waves
associated with the evolution of charged particles (ions and electrons
under quasi-neutrality conditions) emerging when assuming planar propagation
of the particles with all quantities assumed to depend only along the
traveling direction x and the magnetic field is  transverse to the
direction of motion. Remarkably this situation appears to feature both a solitary
wave propagation due to the effective nonlinearity of the $j \times B$
force~\cite{allen4}, but also, as shown below, to enable periodic traveling waves to arise.

The presentation of the paper is as follows: In Section II, we describe the AA model, we 
present our analytical approaches for the identification of the exact solitary wave solutions 
and the connection of the model to the KdV equation. In Section III, we present the results of our
numerical studies. Section IV summarizes our findings, and briefly discusses potential future 
studies, for the AA and other related models.

\section{The model and its analytical consideration}

\subsection{Presentation of the AA model}

The study of Refs.~\cite{aa1,aa2} concerned electrons and ions in a plasma, where the magnetic 
field is in the $z$-direction, and no variations of the pertinent fields along the $y$ or $z$ 
directions were considered. Furthermore, it was assumed that appreciable amounts of energy are 
given to the particles in the waves (e.g., $10$~keV --see Ref.~\cite{aa1}). The applications of 
that work were intended to lie both in the field of fusion research, as well as in the study of 
astrophysical phenomena, such as the solar wind~\cite{solar}.

Let us denote by $U_1$, $U_2$ the velocities along the $x$-direction of two distinct
masses $m_1$ and $m_2$ and the corresponding velocities along the $y$-direction by $V_1$ and $V_2$. 
We express the densities as $n_1, n_2$ and the charges as $e_1$, $e_2$, with $e_1+e_2=0$. 
Then, the force balance equations and Maxwell field equations are:
\begin{eqnarray}
  U_{1,t} + U_1 U_{1,x} &=&\frac{e_1}{m_1} \left[E_x + \frac{V_1 B_z}{c}\right],
                            \label{ueq1}
  \\
  V_{1,t} + U_1 V_{1,x} &=&\frac{e_1}{m_1} \left[E_y - \frac{U_1 B_z}{c}\right],
                            \label{ueq2}
  \\
 U_{2,t} + U_2 U_{2,x} &=&\frac{e_2}{m_2} \left[E_x + \frac{V_2 B_z}{c}\right],
                            \label{ueq3}
  \\
  V_{2,t} + U_2 V_{2,x} &=&\frac{e_2}{m_2} \left[E_y - \frac{U_2 B_z}{c}\right],
                            \label{ueq4}
  \\
  \frac{\partial H_z}{\partial x} &=&-\frac{4 \pi}{c} \left[n_1 e_1 V_1 + n_2
  e_2 V_2 \right],
  \label{eq5}
  \\
  \frac{\partial E_y}{\partial x}&=&-\frac{1}{c} \frac{\partial
  B_z}{\partial t}, 
  \label{eq6}
\end{eqnarray}
where $c$ denotes the speed of light. It is now assumed that the velocities along the 
$x$-direction are equal, i.e., $U_1=U_2$, and similarly the densities for the electrons 
and ions are also equal, i.e., $n_1=n_2$; thus, the relative velocity between them can be 
referred to as $V=V_1-V_2$. A key consideration here is the {\it quasi-neutrality} of the plasma. 
This means that there is a very small difference between $n_1$ and $n_2$ responsible for the 
electric field, yet we may approximately assume $n_1=n_2 \equiv n$ for practical purposes,
an assumption which is valid when the electron plasma frequency is much greater than the 
electron gyrofrequency. The relevant criterion given in Ref.~\cite{aa1} 
also provides the condition for the non-relativistic equations to be valid.

One can then switch to a Lagrangian (moving with the particles) coordinate system and 
adimensionalize over characteristic scales of the magnetic field ${B}_{\star}$ (i.e., 
$B \mapsto B/{B}_{\star}$) and density ${n}_{\star}$ (i.e., $n \mapsto n/{n}_{\star}$). 
In this framework, the speed is measured in units of the characteristic Alfv{\'e}n speed,  
$V_A=\sqrt{{B}_{\star}^2/(4 \pi \mu  {n}_{\star} (m_1+m_2))}$, where $\mu$  stands for 
the magnetic permeability of free space, while the electric field is measured in units of 
${E}_{\star}=V_A {B}_{\star}/c$, i.e., $E \mapsto E/{E}_{\star}$. Finally, the space 
variables $(x,y,z)$ are measured in units of 
$d=\sqrt{m_1 m_2 c^2/(4 \pi n_{\star} \mu e_2^2 (m_1+m_2))}$, while time is
measured in units of ${t}_{\star}=(m_1 m_2)^{1/2} c/(e_2 {B}_{\star})$,
i.e., $t \mapsto t/{t}_{\star}$. As a result, we obtain the dimensionless 
Adlam-Allen (AA) model, described by the following system of PDEs:
\begin{eqnarray}
  R_{tt} &=&-\frac{1}{2} (B^2)_{xx},
             \label{a1}
  \\
  B_{txx} &=& (R B)_t.
  \label{a2}
\end{eqnarray}
Here, $B$ is the rescaled (by ${B}_\star$) magnetic field, while
$R={n}_\star/n_{1,2}$ is the rescaled (inverse) density of ions and electrons. 
Our intention hereafter, is to work with this reduced dimensionless version of the AA model.

\subsection{Fundamental properties of the AA model}
\label{SecA}

\subsubsection{The AA system with vanishing boundary conditions.}

The simplest nontrivial solution of the AA model (\ref{a1})-(\ref{a2}) is expressed in the form:
\begin{equation}
R=R_0, \quad B=B_0,
\label{n0B0}
\end{equation}
where the constants $R_0$ and $B_0$ set the boundary conditions (BCs) at infinity, 
namely $R\rightarrow R_0$ and $B\rightarrow B_0$ as $x\rightarrow \pm \infty$. 
Integrating Eq.~(\ref{a2}) over time, and using the aforementioned BCs, we can 
express the AA model as:
\begin{eqnarray}
R_{tt}&=&-\frac{1}{2}(B^2)_{xx}, \label{aa1} \\
B_{xx}&=&RB-R_0 B_0.
\label{aa2}
\end{eqnarray}
It is now convenient to seek solutions of the AA model on top of the background 
solution, $R=R_0$ and $B=B_0$, namely:
\begin{equation}
R(x,t)=R_0+u(x,t), \quad B(x,t)=B_0+w(x,t), 
\label{nB}
\end{equation}
with the unknown fields $u$ and $w$ satisfying vanishing BCs at infinity, namely 
$u, w \rightarrow 0$ as $x\rightarrow \pm \infty$. Substituting Eq.~(\ref{nB}) 
into Eqs.~(\ref{aa1})-(\ref{aa2}), we obtain the following system of nonlinear PDEs 
for the fields $u$ and $w$:
\begin{eqnarray}
&&u_{tt}+B_0 w_{xx} +\frac{1}{2}(w^2)_{xx}=0,
\label{m1} \\
&&w_{xx}-R_0 w-B_0 u -uw=0.
\label{m2}
\end{eqnarray} 

\subsubsection{Lagrangian structure and integrals of motion.}

It can now be shown that the system of Eqs.~(\ref{m1})-(\ref{m2}) can be derived 
by the Euler-Lagrange equations of fields described by a certain Lagrangian density. 
To derive a Lagrangian for the above system, we follow the methodology used 
in Ref.~\cite{Chalmers} for the derivation of the Lagrangian of the Zakharov's 
equations (see, e.g., Ref.~\cite{ir}), We thus introduce the auxiliary field 
$\rho(x,t)$, such that:
\begin{equation}
u\equiv \rho _{x}.
\label{rho}
\end{equation}%
Inserting Eq.~(\ref{rho}) in Eqs.~(\ref{m1}) and (\ref{m2}), and integrating~(\ref{m1}) 
once with respect to $x$, the AA system is cast in the form:
\begin{eqnarray}
&&\rho_{tt}+B_0 w_x + \frac{1}{2}(w^2)_{x}=0,
\label{au1} \\ 
&&w_{xx}-R_0 w -B_0 \rho_x-\rho_x w=0.
\label{au2}
\end{eqnarray}
Then, it can be found that Lagrangian density corresponding to
Eqs.~(\ref{au1}) and (\ref{au2}) is of the form: 
\begin{equation}
\mathcal{L}=\frac{1}{2}\rho _{t}^{2}+\frac{1}{2}w_{x}^{2}
+\frac{1}{2}R_0 w^2 +\frac{1}{2} \rho_{x}w^{2}+ B_0 \rho_{x} w, \nonumber 
\label{L}
\end{equation}
with the full Lagrangian $L$ defined as the integral of the Lagrangian
density, namely $L=\int_{-\infty }^{+\infty }\mathcal{L}dx$. Indeed, it is 
straightforward to check that the Euler-Lagrange equations:
\begin{equation}
\frac{\partial}{\partial t} \left(\frac{\partial \mathcal{L}}{\partial \Phi_{i,t}}\right)
+ \frac{\partial}{\partial x} \left(\frac{\partial \mathcal{L}}{\partial \Phi_{i,x}}\right)
- \frac{\partial \mathcal{L}}{\partial \Phi_i}=0,\nonumber\
\label{el}
\end{equation}
where $\Phi_i$ (with $i=1,2$) is a generic name for the fields $\rho(x,t)$ and $w(x,t)$ 
respectively, lead to Eqs.~(\ref{au1}) and (\ref{au2}). Furthermore, we can define 
momentum densities $\mathcal{\pi}_i=\partial \mathcal{L}/\partial \Phi_{i,t}$, and 
also introduce the Hamiltonian density $\mathcal{H}$, through the Legendre transformation 
$\mathcal{H}=\sum_i \mathcal{\pi}_i \Phi_{i,t}-\mathcal{L}$. Observing that 
$\mathcal{\pi}_1=\rho_t$ and $\mathcal{\pi}_2=0$, we find that $\mathcal{H}=\rho_t^2-\mathcal{L}$. 
Hence, the system possesses an important conserved quantity, namely the total energy (full Hamiltonian), given by:
\begin{eqnarray}
H \equiv \int_{-\infty }^{+\infty }\mathcal{H}dx 
=\int_{-\infty }^{+\infty } \left(\frac{1}{2}\rho_{t}^{2}-\frac{1}{2}w_{x}^{2}
-\frac{1}{2}R_0 w^2 -\frac{1}{2} \rho_{x}w^{2}- B_0 \rho_{x} w\right)dx.\nonumber
\label{H}
\end{eqnarray}
On the other hand, the system~(\ref{m1})-(\ref{m2}) possesses 
another important conserved quantity, namely the momentum, which is given by:
\begin{equation}
P \equiv \int_{-\infty }^{+\infty }\mathcal{P}dx =-\int_{-\infty }^{+\infty }\rho_t \rho_x dx.\nonumber
\end{equation}
Indeed,  
$dP/dt = \int_{-\infty }^{+\infty } (\rho_{tt} \rho_x+\rho_t \rho_{xt})dx
= \int_{-\infty }^{+\infty }  w_x(R_0 w-w_{xx}) + \frac{1}{2}
(\rho_t^2)_x dx  =0$,
due to the fact that $w$
and $\rho_t$ are assumed to  satisfy vanishing BCs at infinity.
Note that, in the case of traveling wave solutions that we study below, i.e., for 
$u=u(\xi)$ with $\xi=x-vt$, it is straightforward to find that the momentum $P$ is 
reduced to the form:
\begin{equation}
\label{momv}
P=v \int_{-\infty }^{+\infty }u^2dx.
\end{equation}

\subsubsection{The linear regime and connection with the Boussinesq equation.} 

Let us now consider elementary solutions of the AA model, in the form of linear waves,  
close to the background solution~(\ref{n0B0}). To find such solutions, we assume that
$u, w \sim O(\epsilon)$, where $0 <\epsilon  \ll 1$ is a small parameter. Then, at order 
$O(\epsilon)$, the linearization of Eqs.~(\ref{m1})-(\ref{m2}) leads to the following 
linear equations: 
\begin{eqnarray}
&&u_{tt}+B_0 w_{xx} =0,
\label{l1} \\
&&w_{xx}-R_0 w-B_0 u =0.
\label{l2}
\end{eqnarray} 
Obviously, using Eq.~(\ref{l2}), one may substitute $u=(w_{xx}-R_0 w)/B_0$ into Eq.~(\ref{l1}), 
and arrive at the following equation for $w$:
\begin{equation}
w_{tt}-C^2 w_{xx} - \frac{1}{R_0}w_{xxtt}=0,
\label{ibe}
\end{equation}
where
\begin{equation}
C^2 \equiv \frac{B_0^2}{R_0},
\label{ss}
\end{equation}
is the square of the speed of small amplitude linear waves in the
long-wavelength (small $k$) limit; see also Eq.~\eqref{dribe} below.
Equation~(\ref{ibe}) has the form of a linearized {\it improved} Boussinesq equation (iBE). 
Generally, the Bousinesq model is known to describe the evolution of bidirectional shallow 
water waves \cite{mjarecent}, and the dynamics of pulses in nonlinear lattices (through a 
continuous approximation) \cite{Rem}. A crucial difference between the standard Boussinesq 
equation and its iBE variant is that the latter is not prone to unphysical long-wavelength 
instabilities (see, e.g., Ref.~\cite{bogolubsky}). The linearized iBE~(\ref{ibe}) admits 
plane wave solutions, $\sim \exp[i(kx-\omega t)]$, 
with the frequency $\omega$ and wavenumber $k$ obeying the dispersion relation:
\begin{equation}
\omega^2= C^2 k^2 \left(1+\frac{k^2}{R_0}\right)^{-1}.
\label{dribe}
\end{equation}

Considering long waves and weak dispersion, such that $k^2/R_0 \ll 1$, and focusing on the 
case of right-going waves, we may approximate the dispersion relation~(\ref{dribe}) as 
$\omega \approx Ck-(C/2R_0)k^3$, which is the dispersion relation of a Korteweg-deVries 
(KdV) equation. This indicates a strong connection of the AA model with the KdV equation.
Indeed, below we will show that the AA model possesses exact traveling wave solutions, 
in the form of solitary waves, which -- in the small-amplitude limit -- transform into 
the KdV solitons.

\subsection{Derivation of exact traveling waves}

Let us now proceed by seeking solutions of the system~(\ref{m1})-(\ref{m2}) in the form 
of traveling waves, namely:
\begin{equation}
u=u(\xi), \quad w=w(\xi); \quad \xi\equiv x-vt,
\label{tw}
\end{equation}
where $v$ denotes the velocity of the waves. Recalling that $u,w\rightarrow 0$ 
as $x\rightarrow \pm \infty$, Eq.~(\ref{m1}) can readily be integrated twice with respect 
to $\xi$, leading to:
\begin{equation}
u=-\frac{1}{v^2}\left(B_0 w+\frac{1}{2}w^2\right).
\label{eu}
\end{equation} 
Substituting Eq.~(\ref{eu}) into Eq.~(\ref{m2}), we obtain the following nonlinear ODE for 
the field $w$:
\begin{equation}
w''=\frac{B_0^2}{v^2C^2}(v^2-C^2)w-\frac{3B_0}{2v^2}w^2 - \frac{1}{2v^2}w^3,
\label{ode}
\end{equation}
where the primes denote differentiation with respect to $\xi$. 
Equation~(\ref{ode}) can be viewed as an equation of motion of a particle in the presence 
of the effective potential $V(w)$ given by:
\begin{equation}
V(w)=-\frac{B_0^2}{2v^2C^2}(v^2-C^2)w^2+\frac{B_0}{2v^2}w^3 + \frac{1}{8v^2}w^4.
\label{potential}
\end{equation}
The total energy of this dynamical system is
\begin{eqnarray}
\label{TE}
E(w,w')=\frac{1}{2}w'^2+V(w)=E_0,
\end{eqnarray} 
where the constant of integration $E_0$ represents the total initial energy of the effective oscillator which is conserved along its motion. 
\begin{figure}[tbp!]
	\begin{center}
	\begin{tabular}{ccc}	
\includegraphics[scale=0.4]{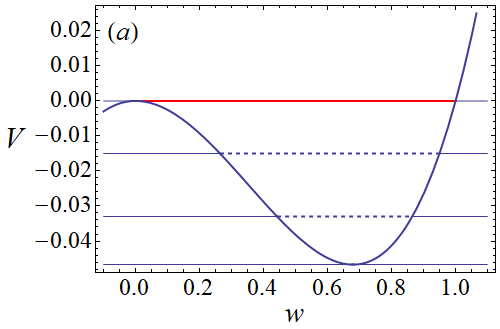}&\hspace{0.3cm} &\includegraphics[scale=0.4]{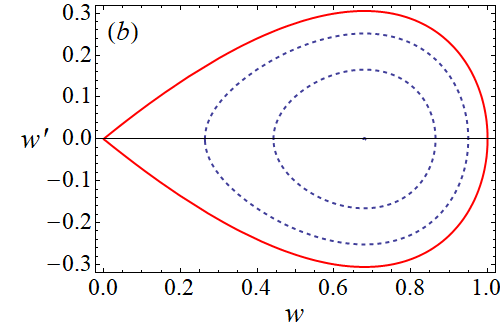}\\[8pt]
\includegraphics[scale=0.4]{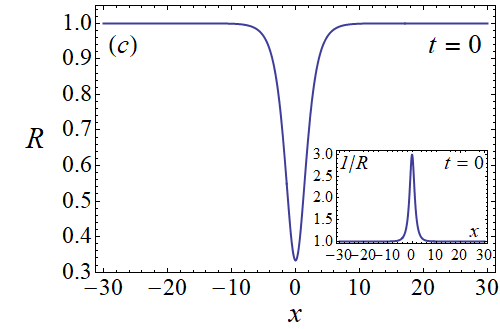}&\hspace{0.3cm} &\includegraphics[scale=0.4]{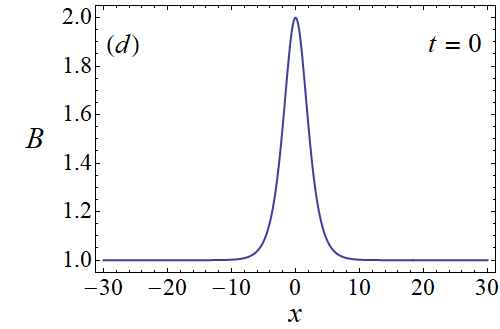}\\[8pt]
\includegraphics[scale=0.4]{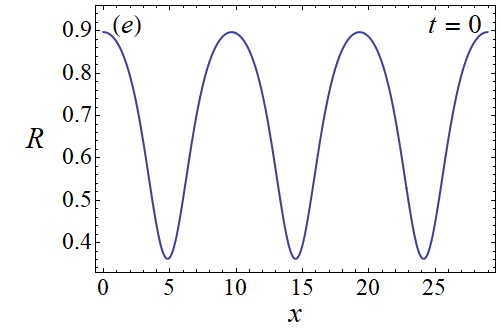}&\hspace{0.3cm} &\includegraphics[scale=0.4]{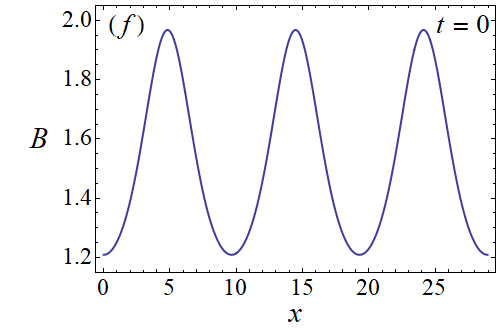}	
	\end{tabular}	
	\end{center}
	\caption{(Color online) Top row: Panel (a) shows the 
effective potential $V(w)$ \eqref{potential}, along with some typical energy levels 
depicted by straight horizontal lines. Parameter values: $B_0=R_0=1$, $v=1.5$. 
The continuous (red) line corresponds to the energy $E_0=V(0)=0$ corresponding 
to the homoclinic orbit. The dashed lines depict energy values $E_0<0$, which correspond 
to closed, periodic orbits. Panel (b) shows orbits in the $(w,w')$ phase-plane of 
Eq.~\eqref{ode}, corresponding to the energy levels of panel (a). The continuous 
(red) curve depicts the homoclinic orbit, pertinent to the energy $E_0=V(0)=0$. 
The dashed closed curves are associated with the periodic solutions corresponding 
to energy values $E_0<0$. Middle row: Panel (c) shows the profile of the 
solitary wave for $R$, given by Eq.~\eqref{nsol}, at $t=0$. Shown also, as an inset, 
is the quantity $1/R$ which, physically, represents the density of the charged particles.  
Panel (d) shows the profile of the soliton solution for $B$ given by \eqref{Bsol}, 
at $t=0$.  Bottom row: a spatially periodic solution corresponding to the energy $E_0=-0.015$. 
Panel (e) depicts the profile of the $R^p$ periodic component of the solution, at $t=0$, while 
the panel (f) depicts the profile of its $B^p$ periodic component, at $t=0$.}
\label{fig:phase_space_soliton}
\end{figure}

A simple analysis shows that if $v<C$, then there exists a sole fixed point at $w=0$ being a 
stable center, at which the potential $V$ attains its global minimum.  
As a result, all solutions of Eq.~(\ref{ode}) are periodic and can, in principle, be expressed in 
terms of the Jacobi elliptic functions. 

On the other hand, if $v>C$ there exist for $w\geq 0$,
two fixed points: an unstable saddle point  at $w=0$ (corresponding to the global maximum of the 
potential $V$ for $w\geq 0$), and a stable center at $w=8(v/C-1)B_0$ (corresponding to the global 
minimum of $V$ for $w\geq 0$). The graph of the potential $V(w)$ \eqref{potential} in this case, is 
portrayed in Fig.~\ref{fig:phase_space_soliton}(a).  This is a case 
of particular interest, since, there exists a homoclinic orbit (separatrix), namely a 
trajectory of infinite period, which corresponds to a solution decaying at infinity, i.e., 
a solitary wave. Figure~\ref{fig:phase_space_soliton}(b) depicts orbits of the 
dynamical system \eqref{ode} corresponding to various energy values (straight horizontal lines shown 
in the graph of the potential $V(w)$, depicted in panel(a)). The homoclinic orbit 
[continuous (red) curve forming a ``loop'' as shown in panel (b)] corresponds to the energy 
$E_0=V(0)=0$ [see the continuous (red) horizontal line shown in the panel (b)].  The closed 
orbits [dashed (blue) curves] correspond to energy values $E_0<0$, and are associated with spatially 
periodic solutions. These solutions will be discussed in Section~\ref{SeC}.

Let us focus on the homoclinic orbit. In order to derive the corresponding solution, 
we use the first integral [see Eq.~(\ref{TE})] for the value $E_0=0$, which is the energy 
of the homoclinic solution. Then, a second integration, 
i.e., $\int dw/\sqrt{-2V(w)}=\xi-x_0$ ($x_0$ is an integration constant), 
leads to the implicit form of the solution, which eventually can be expressed in the following 
explicit form:
\begin{eqnarray}
w(x,t)&=&\frac{2B_0}{C}(v^2-C^2)\frac{1}{C+v\cosh(\theta)}, 
\label{es} \\
\theta &\equiv& \frac{B_0}{vC}\sqrt{v^2-C^2}(x-vt-x_0). 
\label{th}
\end{eqnarray}
In Eq.~\eqref{th}, the integration constant $x_0$ represents the initial position 
of the solitary wave.
Notice that due to the aforementioned necessary condition, $v>C$, for the existence of the 
solitary wave~(\ref{es}), it turns out that this exact traveling wave solution is in fact 
traveling with a speed larger than that of the long-wavelength linear waves.
On the other hand, the constraint for a positive density, $R>0$, yields [using 
Eqs.~\eqref{nB}, \eqref{eu}, and \eqref{es}] an upper bound for the speed, 
namely $v<2C$. Thus, in terms of the velocity $v$, the domain of existence of a
physically relevant solitary wave is:
\begin{eqnarray}
\label{condv}
C<v<2C.
\end{eqnarray}

Summarizing our findings presented in this section, we have derived an exact, 
exponentially localized, travelling wave solution of the AA model, of the form:
\begin{eqnarray}
\label{nsol}
R(x,t)&=&R_0-\frac{1}{v^2}w(x,t)\left[B_0 +\frac{1}{2}w(x,t)\right]:=R^s(x,t), \\
\label{Bsol}
B(x,t)&=&B_0 + w(x,t):=B^s(x,t),
\end{eqnarray}
with $w(x,t)$ given by Eqns.~(\ref{es})-\eqref{th}.  This is the same
wave
as identified in the original~\cite{aa1} work, yet the approach used
herein
will enable us to obtain a considerably wider family of solutions
in what follows.
The profiles of these solutions at $t=0$ are depicted in the middle row of 
Figure~\ref{fig:phase_space_soliton}. Figure~\ref{fig:phase_space_soliton}(c)
shows the profile of $R^s(x,0)$ and Figure~\ref{fig:phase_space_soliton}(d) shows the profile of $B^s(x,0)$
corresponding to a pulse on top of the finite background $B_0=1$. The inset 
of panel (c) also shows $1/R$ which is proportional to the particle density.

\subsection{Connection to the KdV equation}

The derivation of the linearized iBE~\eqref{ibe} that describes the 
linear properties of the model, as discussed in Section~\ref{SecA}, suggests the possibility 
of establishing an asymptotic connection between the AA system and the KdV equation. To do this, 
it is convenient to employ the so-called long-wave approximation \cite{jeffrey}. In particular, 
taking into regard that, for long waves, the dispersion relation~(\ref{dribe}) can be approximated 
as $\omega \approx Ck-(C/2R_0)k^3$, we may assume that the wavenumber $k$ is of the order 
$O(\epsilon^{p})$, where $0<\epsilon \ll 1$ is a formal small parameter, and $p>0$. Notice 
that, as we will see, the choice of $p$ is not important; hence, without loss of generality, 
we choose $p=1/2$. Then, the substitution $k \mapsto \epsilon^{1/2}k$ into the dispersion 
relation leads to the frequency $\omega \approx \epsilon^{1/2}Ck -\epsilon^{3/2}(Ck^3/2R_0)$. 
Accordingly, the phase of plane wave solution of the linearized iBE, Eq.~(\ref{ibe}), 
reads $kx-\omega(k)t \approx \epsilon^{1/2}k(x-Ct)+\epsilon^{3/2}(Ck^3/2R_0)t$. This suggests 
the introduction of the following slow variables:
\begin{equation}
X=\epsilon^{1/2}(x-Ct), \quad T=\epsilon^{3/2}t. 
\label{slow}
\end{equation}
Notice that this choice is consistent with the fact that, in the long-wavelength 
limit of $k \rightarrow 0$, the asymptotic behavior of the solution of the iBE is 
\cite{jeffrey}: $u(x,t) \sim {\rm Ai}(z)$, where 
${\rm Ai}(z)\equiv (1/\pi)\int_0^{+\infty}\cos(sz+\frac{1}{3}s^3) ds$ is the Airy function, 
and $z = (x-Ct)/[(3C/2R_0)^{1/3}t^{1/3}]$. In other words, the asymptotic analysis 
suggests a similarity behavior for a coordinate system with $z={\rm const.}$, which is 
obviously valid for the choice of the coordinate system of Eq.~(\ref{slow}). 

Using the above slow variables, Eqs.~(\ref{m1})-(\ref{m2}) are respectively expressed as follows:
\begin{eqnarray}
&&\epsilon \left(C^2-\frac{B_0^2}{R_0} \right)w_{XX} 
-\epsilon \frac{1}{R_0}\partial_X^2 \left( \frac{1}{2}B_0 w^2 -C^2 uw \right) 
\nonumber \\
&&-\epsilon^2 \left( 2C w_{XT} -\frac{C^2}{R_0}w_{XXXX} \right) =O(\epsilon^j), \quad j\ge 3,
\label{ie1} \\ 
&&\epsilon w_{XX} -R_0 w -B_0 u-uw=0.
\label{ie2}
\end{eqnarray}
It can now readily be observed that the first term in the right-hand side of Eq.~(\ref{ie1}) 
vanishes -- see Eq.~(\ref{ss}). Furthermore, we introduce the following perturbation 
expansions of the fields $u$ and $w$ with respect to $\epsilon$: 
\begin{equation}
u=\epsilon u_1 + \epsilon^2 u_2+ \cdots, 
\quad 
w=\epsilon w_1 + \epsilon^2 w_2+ \cdots,
\label{asex}
\end{equation}
where the powers in $\epsilon$ are chosen so that the dominant dispersion term and 
the dominant nonlinearity term be of the same order; this choice, as we will see, 
gives rise to soliton solutions. Substituting, we can obtain from Eqs.~(\ref{ie1})-(\ref{ie2}) 
the following results. First, at order $O(\epsilon)$, we derive from Eq.~(\ref{ie2}) the 
following equation connecting the unknown fields $u_1$ and $w_1$:
\begin{equation}
u_1=-\frac{R_0}{B_0} w_1.
\label{con}
\end{equation}
Next, using Eq.~(\ref{con}), the nonlinear contribution [second term in Eq.~(\ref{ie2})] arising at 
$O(\epsilon^3)$ in Eq.~(\ref{ie2}) becomes $-(3B_0/R_0)\left(w_1 w_{1X}\right)_X$. As a result, at 
$O(\epsilon^3)$, integration of Eq.~(\ref{ie2}) over $X$ leads to the following KdV equation:
\begin{equation}
2Cw_{1T}+\frac{C^2}{R_0} w_{1XXX} +\frac{3B_0}{R_0}w_1 w_{1X} =0.
\label{kdv}
\end{equation}

It is well known that the KdV equation is a completely integrable system possessing soliton 
solutions (see, e.g., Ref.~\cite{mjarecent}). In particular, the single soliton solution 
of Eq.~(\ref{kdv}), when expressed in terms of the original variables $x$ and $t$, gives 
rise to the following approximate solution [valid up to order $O(\epsilon)$] of the 
system~(\ref{m1})-(\ref{m2}):
\begin{eqnarray}
w(x,t)&=&\epsilon \kappa^2 \frac{4B_0}{R_0}{\rm sech}^2(\eta), 
\label{kdvsol} \\
u(x,t)&=&-4\epsilon \kappa^2 {\rm sech}^2(\eta), 
\label{kdvsol2} \\
\eta &\equiv& \epsilon^{1/2} \kappa \left[x-C\left(1+\epsilon \kappa^2\frac{2}{R_0} 
\right)t-x_0 \right], 
\label{ak}
\end{eqnarray}
where $\kappa$ is an arbitrary $O(1)$ parameter characterizing the amplitude,  the width and  the
velocity of the KdV soliton. Thus, the original AA system~(\ref{a1})-(\ref{a2}) supports 
the following approximate solution:
\begin{eqnarray}
R(x,t)&\approx& R_0 \left[1-\epsilon \frac{4\kappa^2}{R_0} {\rm sech}^2(\eta)\right], \\
B(x,t)&\approx& B_0 \left[1+\epsilon\frac{4 \kappa^2}{R_0}{\rm sech}^2(\eta)\right].
\end{eqnarray}

We will now show that in the limit of $v \rightarrow C$, the solitary wave~(\ref{es}) 
becomes the KdV soliton~(\ref{kdvsol}). To do this, we use $v \approx C$, and approximate 
$v^2-C^2$ as follows: $v^2-C^2 = (v+C)(v-C) \approx 2C(v-C)$. Then, using the identity 
$1+\cosh(\theta)=2\cosh^2(\theta/2)$, we may rewrite Eq.~(\ref{es}) in the form:
\begin{equation}
w(x,t)=\frac{2B_0}{C}(v-C) {\rm sech}^2 \left[\frac{B_0}{2vC}\sqrt{v^2-C^2}(x-vt-x_0)\right].
\label{apso}
\end{equation}
It is now evident that much like the KdV soliton, the amplitude and width of the solitary wave~(\ref{apso}) are set by a single parameter, namely $v-C$. The limit $v\rightarrow C$ 
is, in fact equivalent with the small amplitude limit, whereby $v-C$ plays now the 
role of a small parameter. Then, employing the small parameter $\epsilon$, and noticing that  $\kappa^2 (4B_0/R_0) =O(1)$ as per our analysis, we can set $v-C=\epsilon \kappa^2 C (2/R_0)$.
%
%
This automatically implies that the amplitudes of the solitary wave~(\ref{apso}) and the 
KdV soliton~(\ref{kdvsol}) are equal. Furthermore, it can readily be observed that the 
velocity of the solitary wave~(\ref{apso}) becomes 
$$v=C[1+\epsilon \kappa^2 (2/R_0)],$$
which is equal to the velocity of the KdV soliton~(\ref{kdvsol}).
Finally, the inverse width of the solitary wave~(\ref{apso}) reads:
$$\frac{B_0}{2vC}\sqrt{v^2-C^2} \approx \frac{B_0}{2vC}\sqrt{2C(v-C)} = 
\frac{B_0}{2vC} \sqrt{2C^2 \epsilon \kappa^2 (2/R_0)} = \epsilon^{1/2}\kappa,$$
where we have used the definition of the velocity $C$ in Eq.~(\ref{ss}). 
Thus, the inverse width of the solitary wave~(\ref{apso}) becomes equal to the 
one of the KdV soliton~(\ref{kdvsol}). 
Hence, we have shown that in the limit $v\rightarrow C$, i.e., 
in the small-amplitude limit, the solitary wave~(\ref{apso}) transforms into the KdV 
soliton~(\ref{kdvsol}), which further highlights the asymptotic connection of the 
AA model with the KdV equation.
 
Concluding this section, it is also relevant to make some additional comments. 
The derivation of the KdV equation and the soliton solution relies on the leading-order 
solution [Eq.~(\ref{con})] of Eq.~(\ref{ie2}), which is a singularly perturbed equation. 
In principle, this equation could support boundary layer type solutions; nevertheless, the  
derivation of such solutions demands a consistent treatment of the full system, i.e., 
both Eqs.~(\ref{ie1}) and (\ref{ie2}). This would lead to higher-order corrections to the KdV 
equation and the soliton solution thereof. This is indeed a very interesting problem, but 
a relevant study is beyond the scope of this work.

\subsection{A rational solution}
\label{SecR}
As it was shown in Section \ref{SecA}, the AA system supports the exact, exponentially localized, solitary wave solutions~(\ref{nsol})-\eqref{Bsol}. Nevertheless, the system also possess still another exact, but  
weakly localized, solitary wave solution, which features an algebraic decay as the 
travelling wave coordinate $\xi \rightarrow \pm \infty$. This solution exists in the limiting case 
case where $v=C$. Indeed, in this case, the ODE~(\ref{ode}) reduces to the form:
\begin{equation}
w''=-\frac{3B_0}{2v^2}w^2 - \frac{1}{2v^2}w^3,
\label{ode2}
\end{equation}
where, now, the fixed point $w=0$ becomes an {\it inflection point} for the potential $V(w)$. 
In such a situation, it is possible to identify an exact, algebraically decaying, solution 
of Eq.~(\ref{ode2}), of the following form:
\begin{equation}
w(x,t)=-\frac{4B_0}{1+R_0\xi^2}, \quad \xi=x-Ct.
\end{equation}
This rational waveform, gives rise to the following exact solution of the original 
AA model~~(\ref{a1})-(\ref{a2}):
\begin{eqnarray}
\label{r1}
R(x,t) &=& R_0 \left[1-4\frac{1-n_0\xi^2}{(1+R_0\xi^2)^2} \right]:=R^r(x,t), \\ 
\label{r2}
B(x,t) &=& B_0\left(1-\frac{4}{1+R_0\xi^2} \right):=B^r(x,t).
\end{eqnarray}
This solution, although interesting in its own right from a mathematical point 
of view, suggests an unphysical situation, namely a negative
(inverse) density as it is clear from \eqref{r2}.
\begin{figure}[tbp!]
	\begin{center}
		\includegraphics[scale=0.4]{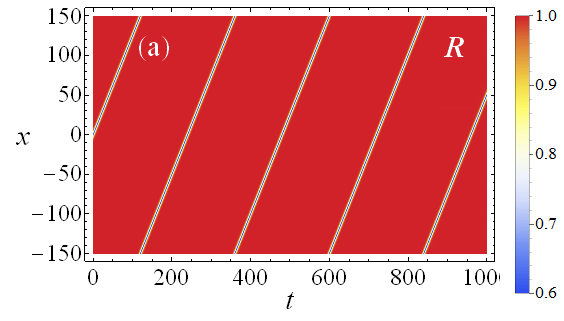}\hspace{0.3cm}
		\includegraphics[scale=0.4]{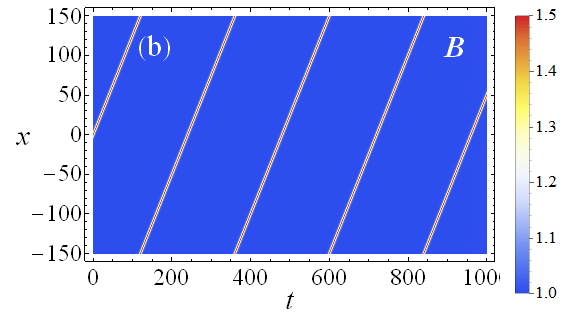}
	\end{center}\vspace{-0.4cm}
	\caption{(Color online) Contour plots showing the spatiotemporal evolution 
	of the solitary waves.
Panel (a) (panel (b)) depicts the evolution of the $R$-component (the
$B$-component)
of the solution. 
Parameter values: velocity $v=1.5$, $B_0=R_0=1$ (i.e., $C=1$), and $L=150$.
	}
	\label{Fig2a}
\end{figure}

\section{Numerical Investigations}

In this section, we perform numerical simulations, aiming to investigate the dynamics of the solitary 
and periodic waves identified in the previous sections. First, we explore the dynamics of the single 
exact solitary wave when considered as an initial condition for the system. Second, we explore the 
dynamics of two solitary waveforms and their potential interaction dynamics. Third, we proceed to a 
numerical study examining the robustness of periodic solutions in the presence of Fourier mode 
perturbations. Finally, due to their mathematical interest (as they are not physically relevant in 
the context of the AA model), we  briefly comment on the rational solutions.

The numerical integration is performed for the following initial-boundary value problem of the AA 
system \eqref{a1}-\eqref{a2},  with initial conditions 
\begin{eqnarray}
\label{ic1}
R(x,0)=\varrho_0(x),\;\;n_t(x,0)=\varrho_1(x),\;\;B(x,0)=\beta_0(x),
\end{eqnarray} 
and periodic boundary conditions on the interval $[-L,L]$, for $B$ and its first derivative $B_x$,
\begin{eqnarray}
\label{bc1}
B(-L,t)=B(L,t),\;\;B_x(-L,t)=B_x(L,t).
\end{eqnarray} 
The initial-boundary value problem \eqref{a1}-\eqref{a2}-\eqref{ic1}-\eqref{bc1} is integrated 
numerically by implementation of the method of lines \cite{MOL};
a tensor product grid discretization scheme is considered for  for the spatial integration, while, for the integration with 
respect to time a 4th–-5th-order adaptive-step Runge-Kutta method 
is used. 

\subsection{Dynamics of the single-soliton solutions}
Numerical integration of the system
\eqref{a1}-\eqref{a2}-\eqref{ic1}-\eqref{bc1} using as initial
conditions the analytical solutions \eqref{nsol}  and \eqref{Bsol} at
$t=0$, i.e.,  $\varrho_0(x)=R^s(x,0)$, $\varrho_1(x)=R^s_t(x,0)$,
$\beta_0(x)=B^s(x,0)$, verified the stability of their time
propagation
(and as a by-product, the accuracy of the numerical method). Their spatiotemporal evolution is depicted in the contour plots of Figure \ref{Fig2a}. Panel (a) portrays the dynamics of the $R$-component of the numerical solution, while panel (b) the dynamics of the $B$-component of the same solution.  The system  is integrated for $L=150$ and for this choice the initial error at the boundaries is far smaller than the accuracy used in the calculations. The parameter values are $B_0=1$, $R_0=1$, $x_0=0$, while the velocity is $v=1.5>C=1$, as $C$ is defined by \eqref{ss}.  The initial data evolve as the exact soliton solutions \eqref{nsol}  and \eqref{Bsol}, preserving their initial profile and speed. 


\subsection{Interaction dynamics of two solitons}

The second numerical experiment investigates the dynamics of two solitary waveforms, 
initialized by a superposition of the analytically derived soliton solutions presented 
in Section \ref{SecA}, i.e., \eqref{nsol} and \eqref{Bsol}.  More precisely, 
we shall consider the dynamics of initial conditions of the form:
\begin{eqnarray}
R(x,0)&=&R_0-\frac{1}{v_1^2}w_1(x,0)\left[B_0 +\frac{1}{2}w_1(x,0)\right]-\frac{1}{v_2^2}w_2(x,0)\left[B_0 +\frac{1}{2}w_2(x,0)\right]:=R^{ds}(x,0), \label{ts1}\\
B(x,0)&=&B_0 + w_1(x,0)+ w_2(x,0):=B^{ds}(x,0),\label{ts2}
\end{eqnarray}
with $\left. w_i(x,t)=w(x,t)\right|_{(x_0=x_i,v=v_i))}$, $i=1,2$, and $w(x,t)$ given by Eq.~(\ref{es}).

The expressions \eqref{ts1}-\eqref{ts2}  describe the interaction between two solitons 
defined by the analytical solutions \eqref{nsol} and \eqref{Bsol}, with velocities $v_1$ 
and $v_2$, respectively. Here, it should be recalled that the permitted and physically relevant velocities  should be such that $v_i\in(C,2C]=(v_{\mathrm{min}},v_{\mathrm{max}}]$.


We consider the parameter values are $B_0=1$ and $R_0=1$ and, thus, $C=1$. 
First, we examine the case where the velocity of the first
soliton $v_1=1.8$ is close to the upper boundary point
$v_{\mathrm{max}}=2$ of the permitted interval $(1,2)$, and the
velocity of the second soliton $v_2=1.2$ is close to the lower
boundary point $v_{\mathrm{min}}=1$. The system
\eqref{a1}-\eqref{a2}-\eqref{ic1}-\eqref{bc1} is integrated for the
initial conditions $\varrho_0(x)=R^{ds}(x,0)$,
$\varrho_1(x)=R^{ds}_t(x,0)$,  $\beta_0(x)=B^{ds}(x,0)$, for  the
above velocities $v_1$ and $v_2$.  The first soliton is initially
located at $x_{1}=-110$ and the second soliton at 
$x_{2}=-70$. This choice ensures that the waves are well separated, 
and in adequate distance from the boundaries in order not to introduce 
any numerical artifacts. The system is integrated for $t\in[0,200]$ 
and $L=150$, which is a sufficient setting for the study of the interaction 
dynamics of the two solitons. 

\begin{figure}[tbp!]
\begin{tabular}{ccc}
\includegraphics[scale=0.32]{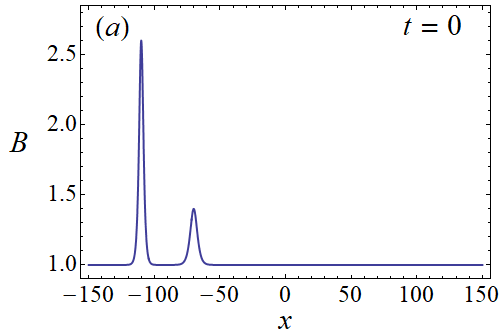}&\includegraphics[scale=0.32]{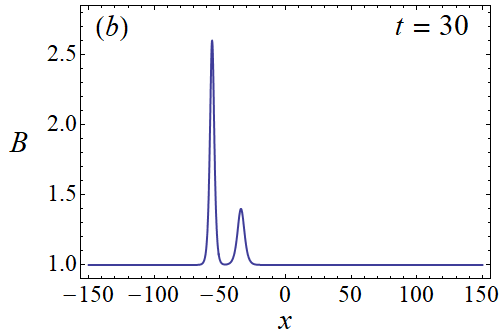}&\includegraphics[scale=0.32]{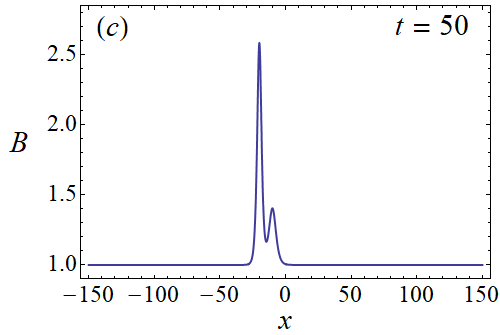}\\
\includegraphics[scale=0.32]{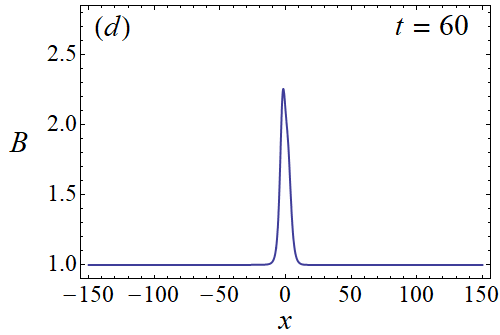}&\includegraphics[scale=0.32]{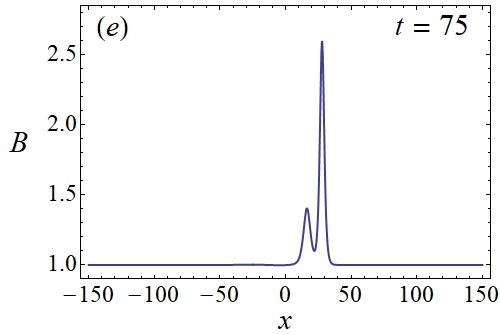}&\includegraphics[scale=0.32]{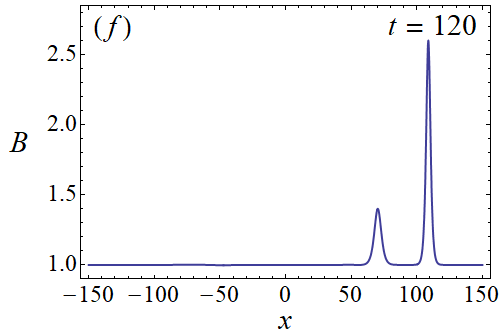}
\end{tabular}
\caption{(Color online) Snapshots of the evolution of two colliding solitary waves.
  In each panel, depicted is the
  $B$-component of the solution. The initial positions and velocities for the solitary 
  waves are $x_{1}=-110$, $v_1=1.8$, and $x_{2}=-70$, $v_2=1.2$. Parameter values are: 
  $B_0=1$, $R_0=1$ and $L=150$.}
\label{snapshots1}
\end{figure}

Figure~\ref{snapshots1} presents snapshots of the evolution of the
$B$-component of the described two-wave initial condition
\eqref{ts1}-\eqref{ts2}, for $t\in [0,120]$.  The snapshots justify
the collision of the two solitons and the near preservation (see the
discussion below) of their initial profiles and velocities after their
collision; the faster and taller soliton of velocity $v_1=1.8$
catches up with the slower and shorter soliton and eventually overtakes it.  

\begin{figure}[tbp!]
\begin{tabular}{ccc}
\includegraphics[scale=0.43]{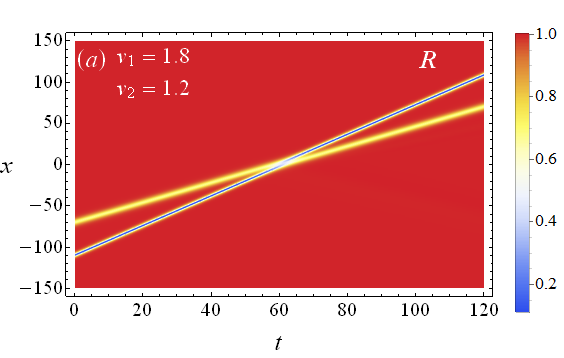}&\hspace{0.15cm}&\includegraphics[scale=0.43]{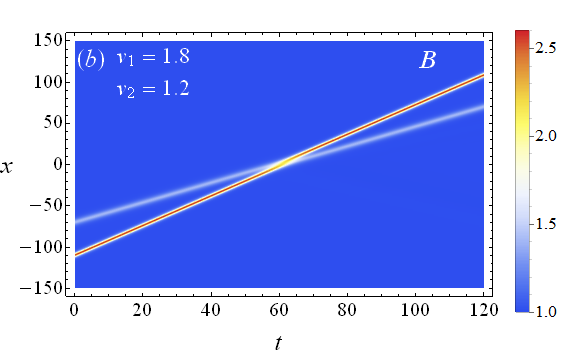}
\end{tabular}
\caption{(Color online) Contour plots showing the spatiotemporal evolution of 
the two-soliton initial conditions $(R^{ds}(x,0), B^{ds}(x,0))$, for velocities $v_1=1.8$ and $v_2=1.2$. The rest of parameters are fixed as in Fig. \ref{snapshots1}. Panel (a) (corr. panel (b)) shows the $R$-component (corr. $B$-component) of the solution. 
}
\label{TwoSolitonsContour1}
\end{figure}

The collision is alternatively visualized by the contour plots of the dynamics 
of the two-soliton initial conditions $R^{ds}(x,0)$ and $B^{ds}(x,0)$, depicted 
in Fig.~\ref{TwoSolitonsContour1}.
Panel (a) depicts the evolution of the $R$-component of the numerical solution, 
while panel (b) depicts the corresponding $B$-component. 

\begin{figure}[tbp!]
\begin{tabular}{ccc}
\includegraphics[scale=0.36]{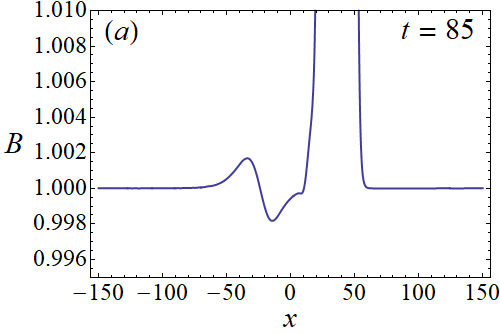}&$\qquad $&\includegraphics[scale=0.36]{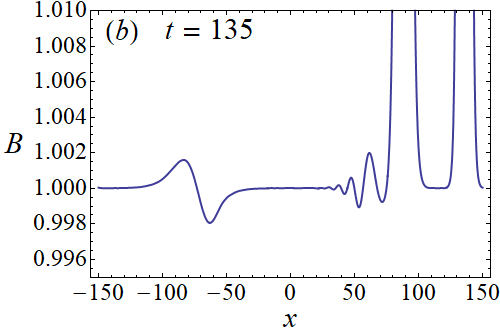}
\end{tabular}
\caption{(Color online) Snapshots of the evolution of the two solitary waves
 with velocities  $v_1=1.8$ and $v_2=1.2$. A magnification around the tail during and after the collision of the solitons, of the $B$-component of the solution is shown. The rest of parameters are fixed as in Fig. \ref{snapshots1}.} 
\label{zoom1}
\end{figure}

An important finding is that the collision of the two solitons is
nearly --but not genuinely-- elastic. This fact suggests the
non-integrability of the system \eqref{a1}-\eqref{a2}. This can be
seen in Figure \ref{zoom1}, depicting snapshots of the dynamics of the
$B$-component of the solution at $t=85$ (panel (a)) and $t=135$
(panel (b)), respectively, for the two solitary wave initial conditions. The snapshots offer a magnified view around the tail of the solution, during and after the collision of the two waves. In both snapshots, we observe the emission of small amplitude wavepackets.

Next, we again fix $B_0=1$ and $R_0=1$, but now we will examine the dynamics of two 
waves initialized through $R^{ds}(x,0)$ and $B^{ds}(x,0)$, with velocities close 
to each other, but also being close to the upper boundary point $v_{\mathrm{max}}=2$ 
of the permitted interval  $(1,2]$. In particular, we use the values $v_1=1.9$ and $v_2=1.8$
and integrate the system for $t\in[0,600]$ and $L=150$. In this case, since the relative 
velocity of the two solitons is smaller, a larger time interval is required in order 
to study the corresponding interaction dynamics.

\begin{figure}[!h]
\begin{tabular}{ccc}
\includegraphics[scale=0.3]{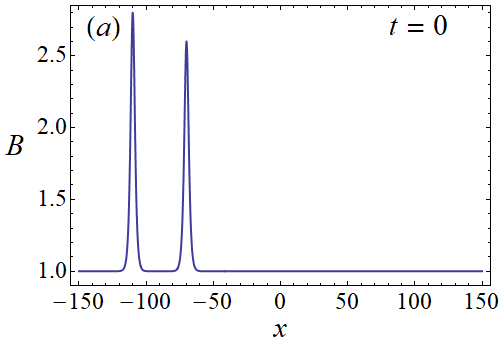}&\includegraphics[scale=0.3]{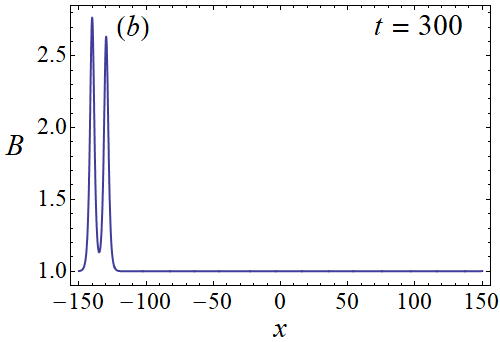}&\includegraphics[scale=0.3]{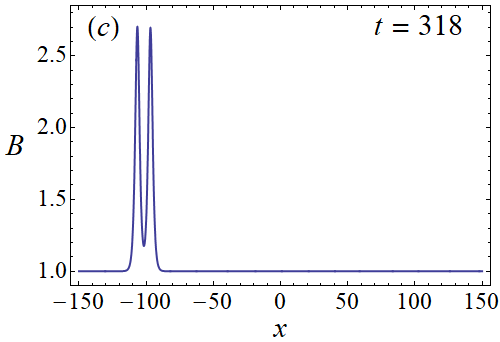}\\
\includegraphics[scale=0.3]{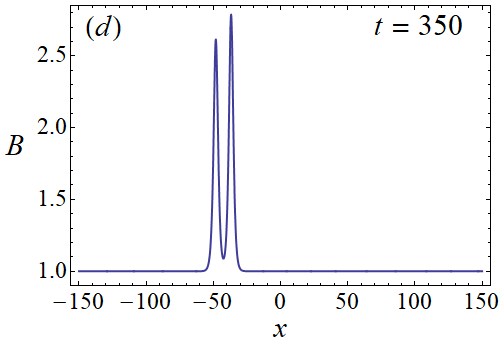}&\includegraphics[scale=0.3]{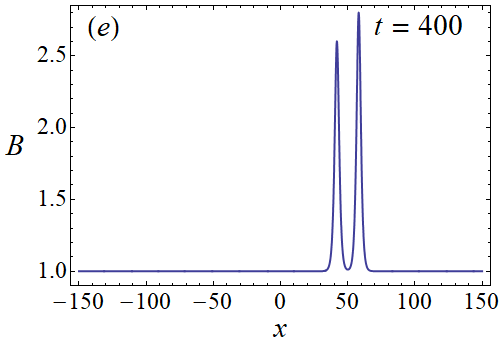}&\includegraphics[scale=0.3]{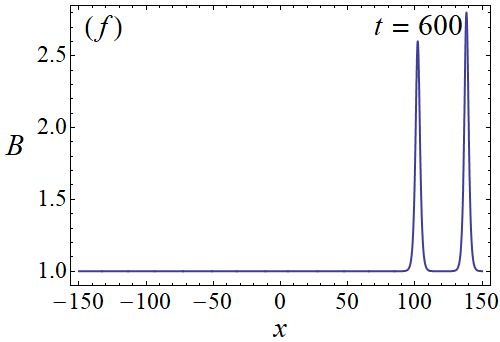}
\end{tabular}
\caption{(Color online) 
Similar to Fig.~\ref{snapshots1}, but now for solitons' initial positions and velocities 
$x_{1}=-110$, $v_1=1.9$ and $x_{2}=-70$, $v_2=1.8$. Parameter values are: $B_0=1$, $R_0=1$ and $L=150$. }
\label{snapshots2}
\end{figure}

\begin{figure}
\begin{tabular}{ccc}
\includegraphics[scale=0.45]{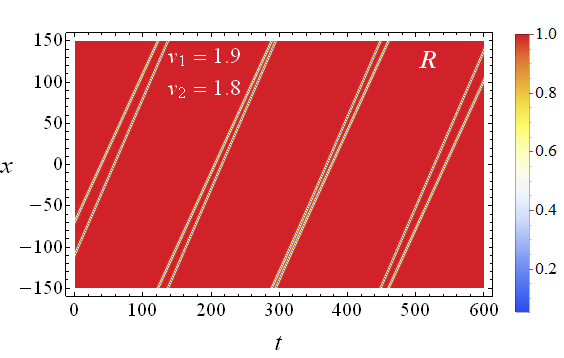}&$\quad $&\includegraphics[scale=0.45]{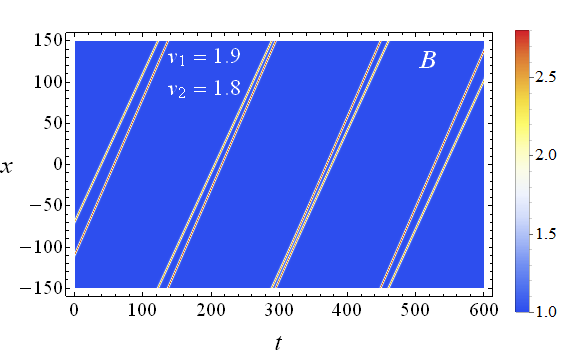}
\end{tabular}
\caption{(Color online) Contour plots of the spatiotemporal evolution
  of the two
  solitary wave initial conditions $(R^{ds}(x,0)$,  $B^{ds}(x,0))$, for $v_1=1.9$ and $v_2=1.8$. The rest of parameter values are fixed as in Fig. \ref{snapshots2}. Panel (a): The $R$-component of the solution is shown. Panel (b): The same as before but for the $B$-component.}
\label{TwoSolitonsContour2}
\end{figure}

Figure~\ref{snapshots2} presents snapshots of the evolution of the
two wave initial condition $(R^{ds}(x,0), B^{ds}(x,0))$; in each panel, 
depicted is the $B$-component of the solution. 
In this case, we observe an interaction of repulsive type: the two
waves exchange their velocities, and afterwards, they continue their motion in 
the same direction. This interaction is also illustrated in the contour plots of 
Fig.~\ref{TwoSolitonsContour2}; panels (a) and (b) show the dynamics of the $R$- 
and $B$-component of the solution, respectively.
\begin{figure}
\begin{center}
\includegraphics[scale=0.3]{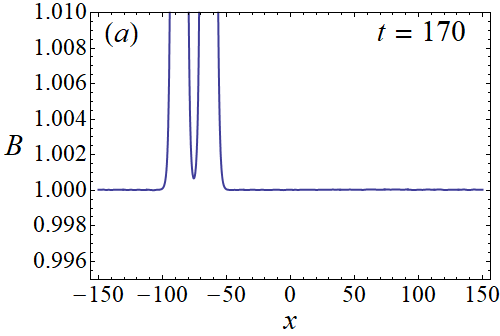} \includegraphics[scale=0.3]{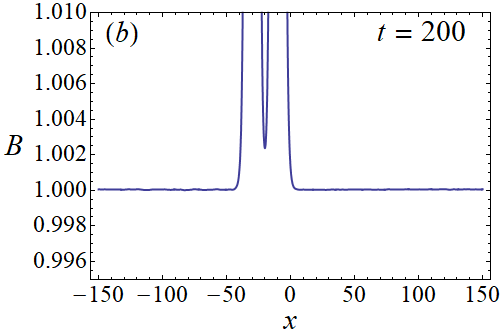} \includegraphics[scale=0.3]{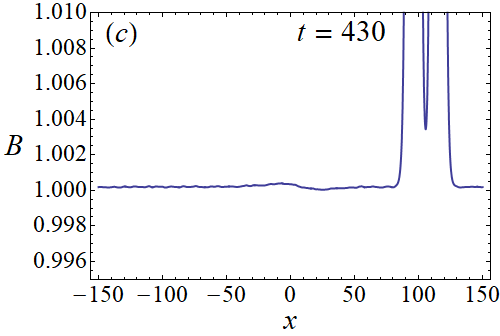}
\end{center}
\caption{Snapshots of the evolution of the two wave initial condition $(R^{ds}(x,0), B^{ds}(x,0))$, with velocities  $v_1=1.9$ and $v_2=1.8$. A magnification around the tail, during the repulsive interaction of the solitons, of the $B$-component of the solution is shown. The rest of parameter values are fixed as in Fig. \ref{snapshots2}.}
\label{zoom2}
\end{figure}

In this case, the emission of
small amplitude waves is weaker if compared with the one of the
previous example for the velocities $v_1=1.8$ and $v_2=1.2$ (where we
observed the collision). This is evident in Fig.~\ref{zoom2},
presenting snapshots of the evolution of the $B$-component of the
solution, with a magnified view around its tail; the emerging
wavepackets are weak yet discernible in the snapshot for $t=430$
(panel (c)). Effectively, it can be observed here that the weak relative kinetic
energy of the structures is not sufficient to overcome the potential energy
barrier of the waves' repulsive interaction. As a result, a minimal 
dispersive wake  is only emitted in the process.
%


\subsection{On the stability of the traveling solitary waves} 

\begin{figure}
\includegraphics[scale=0.45]{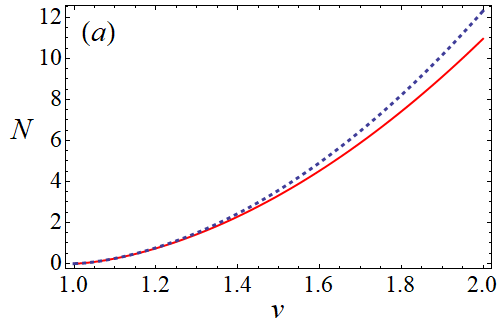}\hspace{0.6cm}
\includegraphics[scale=0.45]{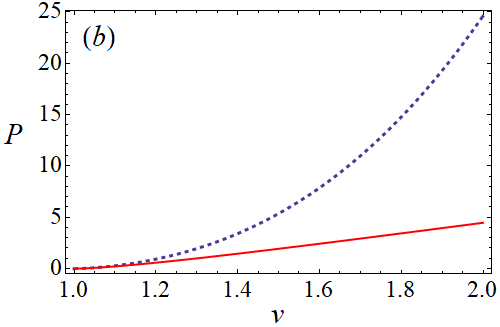}
\caption{(Color online) Panel~(a): The dependence of $N$ on the velocity $v$. 
Here, $N(v)$ has been calculated in the small velocity (KdV) 
limit, using Eq.~\eqref{apso} [dashed (blue) curve], and for all permitted velocities, 
using Eqs.~\eqref{m1}-\eqref{m2} [solid (red) curve]. It is observed that, in both cases, 
$dN/dv>0$, which implies stability of the traveling solitary waves according to the 
KdV stability criterion (see text). 
Panel~(b): The dependence of the momentum $P$ [Eq.~(\ref{momv})] on the velocity $v$. 
As in panel~(a), dashed (blue) curve and solid (red) curve correspond to the KdV limit 
and the exact traveling wave [Eqs.~\eqref{m1}-\eqref{m2}]. Observe that, in both cases, 
$dP/dv>0$ also suggesting stability, per the discussion in the text.
}
\label{dNdv}
\end{figure}

Here, we comment on the stability of traveling solitary waves, relying on established 
stability criteria (see below). We start with the limit of $v\rightarrow C$, 
where the traveling solitary waves reduce to the KdV soliton [see Eq.~(\ref{kdvsol})], 
as shown in the previous Section. In this case, the relevant criterion,
known for numerous decades~\cite{PegoWein92,pelin,kuz1}, states: the KdV soliton is stable if 
$dN/dv>0$, where $N[w]=\int_{-\infty}^{\infty}w^2 dx$ 
%
%
is a conserved quantity -- namely the ``momentum'' -- of the KdV
equation.
Figure~\ref{dNdv}(a) shows the dependence of $N[w]$ on the 
velocity $v$ for the KdV soliton of Eq.~(\ref{apso}) [dashed (blue) curve]. 
In addition, shown also is the dependence of $N[w]$ on the velocity $v$ 
of the exact solitary wave solution of Eq.~(\ref{es}) [solid (red) curve]; 
naturally, the latter curve approaches the former one (in line with
our
reduction) in the limit of $v \rightarrow C$. It is readily seen that
the KdV stability criterion is satisfied, definitively illustrating
that our waves with speeds near the Alfv{\'e}n speed are dynamically stable.
%
%
%

Another interesting aspect worth mentioning involves the recent work
of~\cite{todd} on quadratic operator pencils and associated stability
criteria. Examining the condition of Eq.~(4.8) therein in the case of
a Klein-Gordon model, a short calculation shows that the relevant
quantity
controlling the stability amounts to the derivative of the
Klein-Gordon momentum $P$ as defined by Eq.~(\ref{momv}).
Once again, the positivity of the relevant momentum implies stability
for the Klein-Gordon model, while the negativity thereof implies instability.
Having derived the momentum of the AA system in Sec.~II.2 [see
Eq.~\eqref{momv}],
and given its structural similarity with that of Klein-Gordon models,
we have illustrated its dependence on the speed
in Fig.~\ref{dNdv}(b). Here it is shown that $P$ is an increasing function of $v$ both 
in the KdV limit [dashed (blue) curve] and for the exact solitary wave
solution of the AA model [solid (red) curve].
Naturally, we appreciate that our model is neither of the KdV,
nor of the Klein-Gordon variety directly, hence while these criteria
are suggestive (in that our model has a KdV limit, and a Klein-Gordon
type momentum), it remains an open question to rigorously illustrate
the stability of traveling solitary waves of the present
model. Nevertheless, all of the above observations, as well as our
direct numerical computations for all relevant speeds provide multiple
pointers towards the generic stability of the Adlam-Allen traveling
solitary waves and lead us to conjecture that this feature holds
 in the full range of (physical) 
solitary wave velocities ($C <  v<2C$).




\subsection{Periodic Solutions}
\label{SeC}
\begin{figure}
\begin{tabular}{ccc}
\includegraphics[scale=0.45]{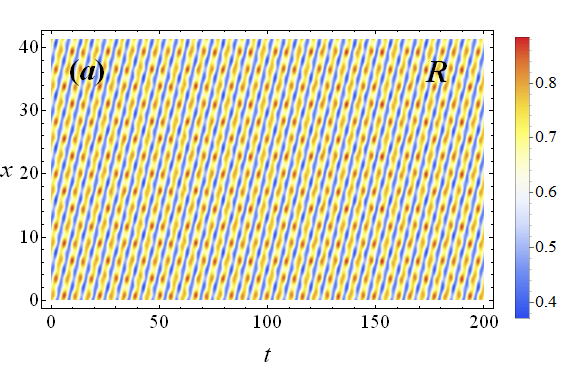}&$\quad $&\includegraphics[scale=0.45]{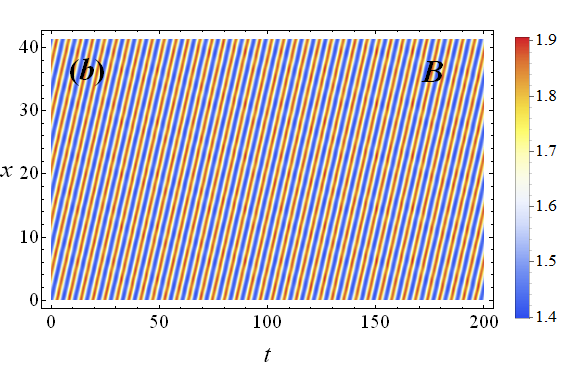}
\end{tabular}
\caption{(Color online) Contour plot of the spatiotemporal evolution of the perturbed spatially periodic solution $(R^p(x,0), B^p(x,0)+0.01\sin(3 K x))$. 
Panel (a): The $R$-component of the solution. Panel (b): The $B$-component of the solution.
Parameter values: $B_0=R_0=1$, $v=1.5$. The spatial interval is
$[0,m\lambda]$ with $m=5$; $\lambda$ is the spatial period of the
solution and $K=2\pi/\lambda$ is the wavenumber associated with the
wavelength $\lambda$. The energy associated with the unperturbed
initial
conditions is $E_0=-0.03$.}
\label{periodic_pert1}
\end{figure}
\begin{figure}
\begin{tabular}{ccc}
\includegraphics[scale=0.45]{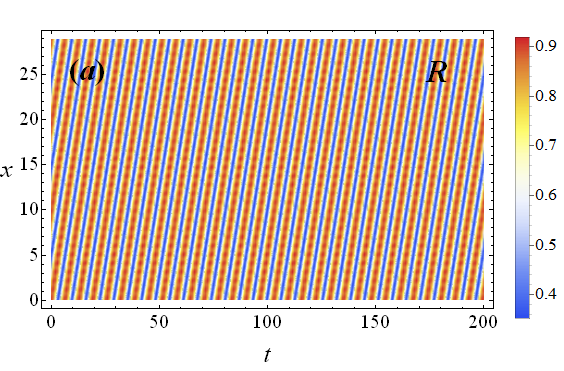}&$\quad $&\includegraphics[scale=0.45]{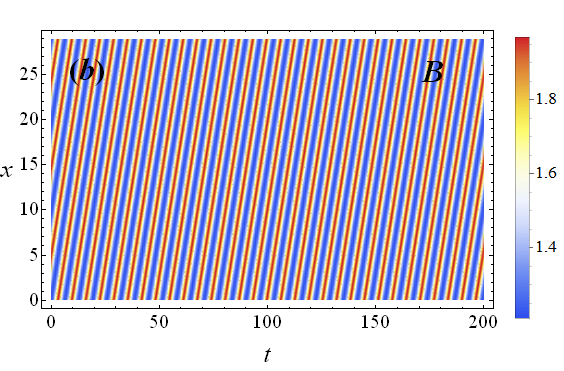}
\end{tabular}
\caption{(Color online) Contour plot of the spatiotemporal evolution of the perturbed spatially periodic solution $R^p(x,0)+0.01\sin(10 K x), B^p(x,0))$ with $m=3$. Panel (a): The $R$-component of the solution. Panel (b): The $B$-component of the solution.
Parameter values: $B_0=R_0=1$, $v=1.5$. Energy of the unperturbed initial conditions $E_0=-0.01$. }
\label{periodic_pert3}
\end{figure}
As was analyzed in Section \ref{SecA}, and visualized in Figure \ref{fig:phase_space_soliton}, for energy values $E_0<0$, we detect spatially periodic, traveling wave solutions, associated with the closed curves inside the homoclinic loop of the $(w,w')$ phase-plane.  Let us recall that the bottom row of Figure \ref{fig:phase_space_soliton} shows the profiles of the $R^{p}(x,t)$-component (panel (a)), and of the $B^{p}(x,t)$-component (panel (b)) at $t=0$, for such a spatially periodic solution (as denoted
by the superscript $^p$), corresponding to the energy  $E_0=-0.015$.
We investigated numerically  the stability of these spatially periodic
solutions in the presence of small amplitude, Fourier mode
perturbations, considered as initial conditions
of the dimensionless problem.
Instead of the symmetric interval $[-L,L]$, the periodic boundary conditions \eqref{bc1} are implemented on the interval $[0,m \lambda]$, where $\lambda$ stands for the wavelength of the solution. The wavelength $\lambda$ can be calculated for a given set of parameters, by integration of Eq. \eqref{TE}. In all of the examples considered herein, we assumed $B_0=R_0=1$, and $v=1.5$.  The results are shown for $t\in [0,200]$, for the sake of clarity of the presented graphics; however, we confirmed that the relevant solutions persist for at least twice the time horizon shown.

In Figure \ref{periodic_pert1}, the spatiotemporal evolution of a perturbed spatially periodic initial condition, is shown. The unperturbed initial condition $(R^p(x,0), B^p(x,0))$ explored is the one with energy $E_0=-0.03$ where we have considered $m=5$. In this condition we perturb the component $B$-component by adding a Fourier mode and the initial condition becomes  $(R^p(x,0), B^{p}(x,0)+0.01\sin(3 K x))$, where $K=2 \pi/\lambda$ denotes the wavenumber associated with the wavelength $\lambda$ of the solution. 
Moreover, Figure \ref{periodic_pert3} depicts the dynamics of another perturbed spatially periodic initial condition of the form $(R^{p}(x,0)+0.01\sin(10 K x), B^p(x,0))$ for the case example of an $m=3$. The unperturbed initial condition corresponds to energy  $E_0=-0.01$.

In both of the above examples, the evolution of the perturbed spatially
periodic initial condition appears to robustly preserve the relevant
waveform without giving rise to any growth modes,
suggesting its dynamical stability. This is in line with what has been
found in the case of cnoidal waves in the KdV equation; see, e.g., the work of~\cite{bernard}. 
\subsection{Rational Solution}
\begin{figure}
\begin{center}
\begin{tabular}{ccc}
\includegraphics[scale=0.4]{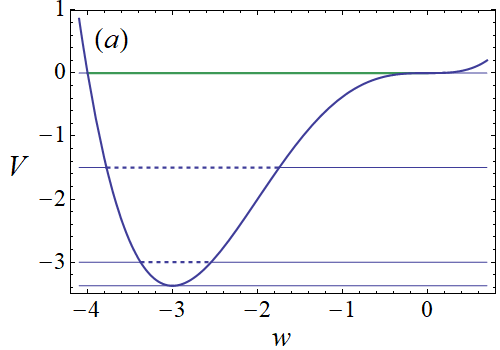}&$\hspace{0.5cm} $&	
\includegraphics[scale=0.4]{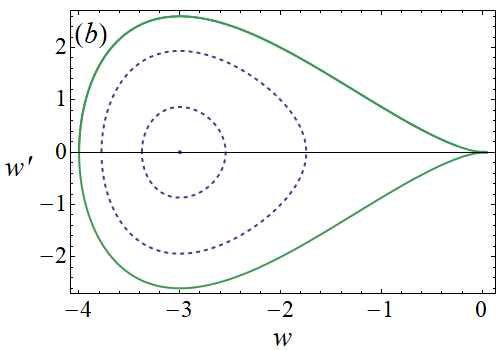}\\
\includegraphics[scale=0.4]{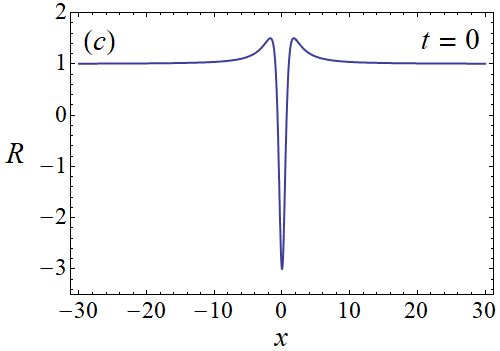}&$ $& \includegraphics[scale=0.4]{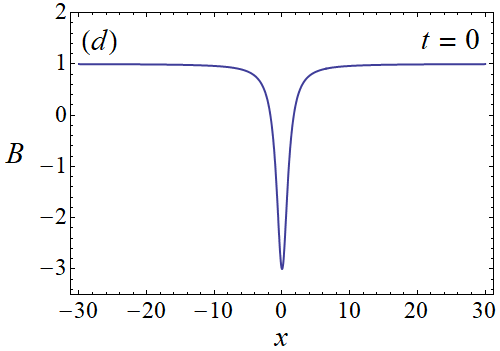}
\end{tabular}
\end{center}
\caption{Panel (a): The graph of the effective potential when $v=C$, the case of the degenerate ODE Eq.~\eqref{ode2}. The (green) horizontal line defines the energy $V(0)=E_0=0$ of the homoclinic connection associated with the analytical rational solutions \eqref{r1}-\eqref{r2}. Panel: (b) The (green) continuous curve is the homoclinic connection of energy $V(0)=E_0=0$. The closed (dashed), periodic orbits correspond to energy values $E_0<0$ (dashed horizontal lines panel (a)). Panel (c): The profile of the rational solution $R^r(x,t)$, given by Eq.~\eqref{r1}, at $t=0$. Panel (d): The profile of the rational solution $B^r(x,t)$, given by Eq.~\eqref{r2}, at $t=0$. }
\label{rational}
\end{figure}

As was shown in Section \ref{SecR}, the case $v=C$ gives rise to a
degenerate
scenario for the  dynamical system \eqref{ode2}. Figure~\ref{rational}(a) shows the graph of the effective potential in this case: the potential $V$ has an inflection point 
at $w=0$, and the energy  $V(0)=E_0=0$ [continuous (green) horizontal line] defines the energy of a cusp-like homoclinic connection in the $(w,w')$ phase-plane; it is depicted by the continuous (green) curve in panel (b) (see also \cite{Hale}).  The corresponding analytical rational solutions $R^r(x,t)$ and $B^r(x,t)$ of the system are given by \eqref{r1} and \eqref{r2}. The energy values $E_0<0$ (dashed horizontal curves in the panel (a)) are associated with the periodic orbits  in the phase plane (dashed closed curves in panel (b)).

The spatial profiles of the coherent structure which is associated with the rational solution are illustrated in the bottom panels of Figure~\ref{rational}. Panel (c) shows the $R^r(x,t)$ component at $t=0$, and panel (d) shows the $B^r(x,t)$-component, at $t=0$.

We have attempted to integrate the system with initial conditions $R^r(x,0)$ and $B^r(x,0)$, 
but have found that the dynamics is extremely sensitive to small
perturbations giving rise to numerical instabilities. We have noted
also that the kind of numerical instability appearing is dependent on
the choice of the half-length $L$. This is to be expected given the
algebraic decay-rate of the initial conditions. Thus, even for a
choice of $L=1000$ the initial error at the boundaries is of the order
of $\mathcal{O}(10^{-6})$ which can introduce numerical
problems. In particular, considering these as small perturbations,
the dynamical behavior of the solution turns out to be highly unstable.

This is a strong indication of the dynamical instability of this state. Thus, due also to its featuring negative density
in part of the domain (hence being of rather limited physical relevance),
we have not pursued it further. 

\section{Conclusions}
We studied the Adlam-Allen (AA) system of partial differential equations, one of 
the prototypical and fundamental models for the description of hydrodynamic disturbances 
in collisionless plasmas. The phase-plane analysis for the relevant, effective second-order 
conservative dynamical system (associated with the description of traveling waves), 
enabled us to identify exact soliton solutions for the original system.
In line with the original work of~\cite{aa1}, we have found  
that these waves
have velocities between once and twice that of the characteristic Alfv{\'e}n speed. 

Another important finding was that the Adlam-Allen system is strongly 
connected to the Korteweg-de Vries (KdV) equation: we have shown that, 
in the small-amplitude limit, the solitary
waves of the original AA system transform into the soliton solutions
of the  Korteweg-de Vries equation. This connection was already
highlighted, when studying the linearization of the Adlam-Allen system: we found 
that this can be described by a linearized improved Boussinesq equation, which, 
in the long wave approximation and in the weak dispersion regime, features the same dispersion relation as the  Korteweg-de Vries equation. 

The above justifications motivated us to study by direct numerical simulations, 
not only the dynamics of the exact soliton solutions, but also the interaction dynamics 
of two soliton waveforms, initialized through a superposition of the analytical solitary waves. 
First, the stable evolution of individual exact pulses was observed in agreement with the analytical 
arguments for their derivation.
In the more interesting case of  soliton interactions, we examined two
scenarios. In the first scenario where the velocity of the one soliton
is close to the Alfv{\'e}n speed, and the velocity of the second is
considerably higher, we found a quasi-elastic collision: the fast
soliton overtakes the slow-one, and both almost preserve their
velocities and shape. The weak inelasticity of the collision was detected
by the emission of small amplitude linear waves. The occurrence of the latter is further suggesting 
that the Adlam-Allen system might be non-integrable. In the second scenario, where both soliton 
velocities take values close to twice of the Alfv{\'e}n speed, we observed an interaction of 
repulsive type; after an exchange of their velocities, the solitary waves continued their propagation 
in the same direction, emitting far weaker wavepackets (in comparison to the previous case).  
Concerning the stability of the presented traveling solitary waves, we
have brought to bear stability criteria both from the realm of KdV equations,
as well as from that of Klein-Gordon models. Pertinent results 
suggest that the solitary waves are stable (definitively so in the
vicinity of the characteristic Alfv{\'e}n speed), a conjecture that is
also generically supported 
by our direct simulations that were performed in the full range of permitted velocities.
%

The dynamical systems analysis verified also the existence of spatially periodic solutions. A 
numerical study examining the dynamics of such solutions in the presence of small perturbations, 
suggests that these spatially periodic travelling waves might be robust as well. 
Finally, the same dynamical analysis was used to reveal
the existence of rational solutions, possessing an algeraic decaying rate in 
the limiting case of propagation at the Alfv{\'e}n speed. 
It is worthile to note that such solutions  are of growing interest due to their argued relevance within the context of rogue waves \cite{k2b,k2d}.
While such solutions are not of physical interest here (since they feature negative densities in part of the spatial domain), they are certainly of interest from a mathematical point of view. 
Remarkably, they are associated with a degenerate case of the effective conservative dynamical system. Yet, they are unfortunately found to be 
quite unstable numerically.

This study is only a first step towards an attempted revival of the interest in the 
Adlam-Allen model. While we have explored special solutions and their full PDE dynamics, 
numerous questions remain unexplored. Among them, 
from a mathematical analysis point of view, well-posedness
properties (local and global) appear to us worthwhile to study, in analogy to the 
well-established counterpart of the KdV model. A more definitive view of the 
potential integrability of the problem (or, more likely, lack thereof) could be an 
interesting direction to pursue in its own right. Finally, we note that this model 
concerns the analysis of a transverse magnetic field,
while recently~\cite{Abbas19}, the longitudinal, far more complex case
has also been
considered. Expanding the lines of thinking
of the present work regarding multiple solitary waves, and their interactions, 
as well as generalized periodic solutions is also
of interest. Work along these directions is currently in progress and will be reported in future studies.


\begin{thebibliography}{99}

\bibitem{aa1} J. H. Adlam and J. E. Allen, Philosophical Magazine {\bf 3}, 448--455 (1958).

\bibitem{aa2} J. H. Adlam and J. E. Allen, Proc. Phys. Soc. {\bf 75}, 640 (1960).

\bibitem{kz} N. J. Zabusky and M. D. Kruskal, Phys. Rev. Lett. {\bf 15}, 240 (1965).

\bibitem{fpu} E. Fermi, J. Pasta, and S. Ulam, 
 Tech. Rep. Los Alamos Nat. Lab. LA1940 (1955).

\bibitem{kdv} D. J. Korteweg and G. de Vries, Phil. Mag. {\bf 39}, 422 (1895).

\bibitem{tan} H. Washimi and T. Taniuti, Phys. Rev. Lett. {\bf 17}, 996 (1966).

\bibitem{ir} E. Infeld and G. Rowlands, {\it Nonlinear Waves, Solitons and Chaos}  
(Cambridge University Press Cambridge, 1990).

\bibitem{kono} M. Kono and M. M. Skori\'c, {\it Nonlinear Physics of Plasmas}
(Springer-Verlag, Heidelberg 2010).
%
\bibitem{allen3} J. E. Allen, Phys. Scripta {\bf 57}, 436 (1998).

\bibitem{allen4} J. E. Allen, and J. Gibson, Phys. Plasmas {\bf 24}, 042106 (2017).

\bibitem{mjarecent} M. J. Ablowitz, {\it Nonlinear Dispersive Waves: Asymptotic Analysis
and Solitons} (Cambridge University Press, Cambridge, 2011).
%

\bibitem{vakh} N. G. Vakhitov and A. A. Kolokolov, 
Radiophys. Quantum Electron. {\bf 16}, 783 (1973). 

\bibitem{kuz2} E. A. Kuznetsov, Phys. Lett. A {\bf 382}, 314 (1984).

\bibitem{PegoWein92} R. Pego and M. Weinstein, Philos. Trans. Roy. Soc. London Ser. A 
\textbf{340}, 
47
(1992).
%

\bibitem{pelin} D.E. Pelinovsky in:
  {\it Nonlinear physical systems: spectral analysis, stability and bifurcations},
  E.N. Kirillov, D.E. Pelinovsky (Eds.), John Wiley  (Hoboken, 2014),
  p. 377


\bibitem{kuz1} E. A. Kuznetsov, Phys. Lett. A {\bf 1012}, 2049 (2018).

\bibitem{todd} J. Bronski, M.A. Johnson, T. Kapitula,
  Comm. Math. Phys. {\bf 327}, 521 (2014). 



\bibitem{Earth} D. A. Tidman and N. A. Krall, {\em Shock Waves in Collisionless Plasmas} (John Wiley and Sons, 1971).
%
\bibitem{Geophys} K. Sauer, E. Dubinin and J. F. McKenzie, Geophys. Res.Lett. \textbf{29}, 2226 (2002).
%
\bibitem{Plasma} C. M. C.Nairn, R. Bingham and J. E. Allen, J.Plasma Phys. \textbf{71}, 631 (2005).
%
\bibitem{solar} N. Meyer-Vernet, {\it Basics of the solar wind},
  (Cambridge University Press, Cambridge, 2007). 
 %
\bibitem{Chalmers} B. Malomed, D. Anderson, M. Lisak, M. L.
Quiroga-Teixeiro, and L. Stenflo,  Phys. Rev. E \textbf{55}, 962-968 (1997).
%
\bibitem{grad} I. S. Gradshteyn, I.M.  Ryzhik, {\it Table of Integrals, Series, and Products}, 
Academic Press (Cambridge, 1994).

\bibitem{Rem} M. Remoissenet, {\it Waves Called Solitons} (Springer, Berlin, 1999).

\bibitem{bogolubsky} I. L. Bogolubsky, Some examples of inelastic soliton interactions, 
Comp. Phys. Comm. {\bf 13}, 149-155 (1977).

\bibitem{jeffrey} A. Jeffrey and T. Kawahara, 
{\it Asymptotic Methods in Nonlinear Wave Theory} (Boston, MA, Pitman, 1982).

\bibitem{MOL} W. E. Schiesser, {\em The numerical method of lines} (Academic Press, 1991).
%
%
\bibitem{k2b} C. Kharif, E. Pelinovsky, and A. Slunyaev, {\it Rogue Waves in the Ocean} (Springer, New York, 2009).

%
\bibitem{k2d} M. Onorato, S. Residori and F. Baronio, {\it Rogue and Shock Waves in Nonlinear Dispersive Media} (Springer-Verlag, Heidelberg, 2016).

%
\bibitem{bernard} N. Bottman, B. Deconinck,
  Discr. Cont. Dyn. Sys. A {\bf 25}, 1163 (2009).

%
\bibitem{Hale} J. K. Hale and H. Ko\c{c}ak, {\it Dynamics and
    Bifurcations} (Springer-Verlag, New-York,  1991)

%
\bibitem{Abbas19} G. Abbas, J.E. Allen, M. Coppins, L. Simons,
  L. James,
  {\it A study of the propagation of a solitary wave along the
    magnetic
    field in a cold collision-free plasma}, preprint (2019).

\end{thebibliography}
\end{document}